\documentclass[twocolumn,showpacs,pra,aps]{revtex4}
\usepackage{amsmath}
\usepackage{amsfonts}
\usepackage{amssymb}
\usepackage{graphicx}
\usepackage{hyperref}
\usepackage{bm}

\everymath{\displaystyle} 
\newcommand{\pr}[1]{\ensuremath{\left[#1\right]}} 
\newcommand{\pc}[1]{\ensuremath{\left(#1\right)}}

\begin{document}

\title{Shielding vacuum fluctuations with graphene}
\author{Sofia Ribeiro}
\affiliation{Quantum Optics and Laser Science, Blackett Laboratory,
Imperial College London, Prince Consort Road, London SW7 2BW, United Kingdom}
\author{Stefan Scheel}
\affiliation{Institut f\"ur Physik, Universit\"at Rostock, Universit\"atsplatz
3, D-18051 Rostock, Germany}

\date{\today}

\begin{abstract}
The Casimir-Polder interaction of ground-state and excited atoms with graphene
is investigated with the aim to establish whether graphene systems can be used
as a shield for vacuum fluctuations of an underlying substrate. 
We calculate the zero-temperature Casimir-Polder potential from the reflection
coefficients of graphene within the framework of the Dirac model.
For both doped and undoped graphene we show limits at which graphene could be
used effectively as a shield. Additional results are given for AB-stacked
bilayer graphene.
\end{abstract}


\maketitle

\section{Introduction}

Graphene's extraordinary electronic and optical properties hold great promise
for applications in photonics and optoelectronics. The existence of a true
two-dimensional (2D) material having a thickness of a single atom was believed
to be impossible for a long time, because both finite temperature and quantum
fluctuations would destroy the 2D structure. However, since the first
groundbreaking experiments \cite{Science306_666_2004}, the study of graphene
became an active field in condensed matter. Theoretical reviews of graphene's
properties can be found in Refs.~\cite{RMP_81_2009,RMP_Peres_2010}. 
The technological push towards miniaturization resulted in the idea of devising
small structures based on graphene. For instance, placing graphene between
different substrates or by patterning a given substrate, it is possible to
create artificial materials with tunable properties \cite{RPP74_082501_2011}. 

Hybrid quantum systems which combine cold atoms with solid structures hold great
promise for the study of fundamental science, creating the possibility to built
devices to measure precisely gravitational, electric and magnetic fields
\cite{NJP13_083020_2011}. For instance, many of the proposed extensions to the
Standard Model of particle physics include forces, due to compactified extra
dimensions, that would modify Newtonian gravity on submicrometer scales
\cite{BookNonNewton,ARNP53_77_2003}. 
By performing extremely careful force measurements near
surfaces, it is hoped that more stringent limits on the presence of such forces
may be obtained. With this in mind, hybrid systems in which neutral atoms and
graphene are held in close proximity represent an important and attractive case
to study. A quick estimate shows that the Casimir-Polder force dominates gravity
by several orders of magnitude at micrometer distances. It is therefore
necessary to find a system that is simple enough in order to either be able to
calculate its dispersion effect to high enough precision, or to provide a shield
against vacuum fluctuations of another (macroscopic) body.

Graphene has been shown to be a strong absorber of electromagnetic radiation, it
interacts strongly with light over a wide wavelength range, particularly in the
far infrared and terahertz parts of the spectrum due to its high carrier
mobility and conductivity \cite{PRL108_047401_2012}. Considering that graphene
is only one atomic layer thick, its (universal) absorption coefficient of 
$\eta=\pi e^2 / (\hbar c) \approx 2.3\%$ is quite remarkable
\cite{graphenebook}.
In Ref.~\cite{nnano7_330_2012} new systems made of several layers of
graphene are shown to be an effective shield for terahertz radiation, while
letting visible light pass. These studies brought the attention to the
development of transparent mid- and far-infrared photonic devices. 

With graphene's absorption properties in mind we investigate the possibility of
shielding electromagnetic vacuum fluctuations of a macroscopic body placed
nearby. The purpose of this study is to investigate whether and under which
circumstances the Casimir-Polder potential between an atom and a
graphene-substrate system is dominated by the interaction with graphene such
that the effect of the substrate does not play an important role. This knowledge
will allow us to manipulate the Casimir-Polder potential of a layered system by
placing the graphene at different graphene-substrate distances or by patterning 
it into different shapes.

This article is organised as follows. After briefly introducing graphene into
the formalism of macroscopic QED in Sec.~\ref{sec:CPpotentialgraphene}, we
give some numerical results for the Casimir--Polder shift of an atom near a
graphene sheet in Sec.~\ref{sec:atomgraphene}. In Sec.~\ref{sec:1sheet} and
\ref{sec:bilayer}, we study different the shielding of vacuum fluctuations by
single-layer and bilayer graphene, respectively, and give concluding remarks
in Sec.~\ref{sec:conclusions}.

\section{Casimir-Polder interaction with graphene}
\label{sec:CPpotentialgraphene}

It is well known that an atom placed near a macroscopic body will experience a
dispersion force --- the Casimir-Polder force  --- due to the presence of
fluctuations of the electromagnetic field even at zero temperature
\cite{casimirpolderpaper}. We begin by investigating the Casimir-Polder
interaction of an atom next to a graphene layer at zero temperature. We adopt
the Dirac model for graphene and calculate the Casimir-Polder interactions based
on the formalism of macroscopic QED. Upon quantization of the electromagnetic
field in the presence of absorbing bodies, and application of second-order
perturbation theory, the Casimir-Polder potential for planar structures can be
written as \cite{acta2008}
\begin{gather}
U_{\mathrm{CP}} \pc{z_{A}} = \frac{\hbar \mu_{0}}{8 \pi^{2}} \int_{0}^{\infty}
d \xi \xi^{2} \alpha_{at} \pc{i \xi} \nonumber \\
\times \int\limits_{0}^{\infty} d k_{\parallel} \frac{e^{-2 k_{\parallel}
\gamma_{0z} z_{A}} }{\gamma_{0z}} \left[
\mathrm{R}_{\mathrm{TE}} + \mathrm{R}_{\mathrm{TM}} \left( 1- \frac{2
k_{\parallel}^{2} \gamma_{0z}^{2} c^{2} }{\xi^{2}} \right) \right]    
\label{eq:Ucp_1}
\end{gather}
where $\gamma_{iz}=\sqrt{1+\varepsilon_i(i\xi)\xi^{2}k_\|^{2}/c^2}$ is the
$z$-component of the wavenumber in the medium with permittivity $\varepsilon_i$
for imaginary frequencies (the index 0 refers to the medium in which the atom is
placed) and $\alpha_{at} (\omega)$ is the isotropic atomic polarizability
defined by
\begin{eqnarray}
\mathbf{\alpha}_{at} (\omega) = \lim_{\varepsilon \rightarrow 0} \frac{2}{\hbar}
\sum_{k_{A} \neq 0_{A}} \frac{\omega_{k 0} \mathbf{d}_{0 k}\cdot \mathbf{d}_{k 0} }{\omega_{k
0}^{2} -\omega^{2} - i \omega \varepsilon } .
\label{eq:atomicpol}
\end{eqnarray}
This equation is valid for zero temperature. A replacement of the frequency 
integral by a Matsubara sum has to be performed for finite temperatures
\cite{acta2008}. In this case, the potential is well approximated by inserting 
the temperature-dependent reflection coefficients in the lowest term in the 
Matsubara sum ($j=0$) while keeping the zero-temperature 
coefficients for all higher Matsubara terms \cite{PRB84_035446_2011}.
Only for $k_B T \gtrsim \Delta$ thermal corrections become important 
\cite{PRA86_012515_2012} ($\Delta \approx 0.1$~eV is the gap parameter of 
quasiparticle excitations).

In order to compute the reflection coefficients $\mathrm{R}_{\mathrm{TM}}$ and
$\mathrm{R}_{\mathrm{TE}}$ for transverse magnetic (TM) and transverse electric
(TE) waves from a graphene sheet, two alternative models, the hydrodynamic
model and the Dirac model, exist.
The first treats graphene as an infinitesimally thin positively charged flat
sheet, carrying a homogeneous fluid with some mass and negative charge
densities. On the other hand, the Dirac model incorporates
the conical electron dispersion relation of Dirac fermions, modelling graphene
as a two-dimensional gas of massless Dirac fermions, where low-energy
excitations are considered Dirac fermions that move with a Fermi velocity. The
ranges of validity of these respective models is not completely resolved
\cite{NJP13_083020_2011}. 

The reflection coefficients are calculated by matching the dyadic Green
function of free space and its derivatives on either side of a two-dimensional
conducting sheet \cite{PRB85_195427_2012}, with the result that
\begin{eqnarray}
\mathrm{R}_{\mathrm{TM}} &=& \frac{\gamma_{0z} \alpha_{\parallel}
(k_{\parallel}, \omega)}{1+\gamma_{0z} \alpha_{\parallel} (k_{\parallel},
\omega)} ,\nonumber\\
 \mathrm{R}_{\mathrm{TE}} &=& \frac{(\omega / c k_{\parallel})^{2} 
\alpha_{\perp} (k_{\parallel}, \omega) }{\gamma_{0z} -(\omega / c
k_{\parallel})^{2}  \alpha_{\perp} (k_{\parallel}, \omega)},
 \label{eq:rBo}
\end{eqnarray}
where
\begin{equation}
\alpha(\mathbf{k},\omega)=- e^2
\frac{\chi(\mathbf{k},\omega)}{ 2 \varepsilon_0 k_{\parallel}} =  i
\frac{\sigma (\mathbf{k},\omega) \,  k_{\parallel}}{2 \varepsilon_0 \omega}
\end{equation}
is given by the density-density correlation function $\chi(\mathbf{k},\omega)$ 
\cite{NJP8_318_2006, PRB75_205418_2007} or, alternatively, the conductivity
$\sigma (\mathbf{k},\omega)$ \cite{PRB77_155409_2008, PRB86_075439_2012}.
The functions for doped and undoped graphene are derived based on the band
structure of graphene. The problem with this approach is that there are no
transverse functions available in the literature for single sheets, only the
longitudinal functions \cite{PRB85_195427_2012}. It has been shown in 
Ref.~\cite{PRB80_245424_2009} that, at zero as well as at finite temperature,
the retardation effects in graphene systems are negligible. Hence, we neglect
the retardation effects, so that the TE modes do not contribute and we can
set $\gamma_{0z} \to 1$.

The density-density correlation function for undoped graphene in the complex 
frequency plane ($\omega=i\xi$) is given by \cite{PRB85_195427_2012} 
\begin{equation}
\chi (\mathbf{k}, i\xi) = -\frac{g}{16 \hbar}
\frac{k^{2}_{\parallel}}{\sqrt{v_{F}^{2} k^{2}_{\parallel} + \xi^2 }}.
\label{eq:xi_undoped}
\end{equation}
The parameter $g=4$ represents the degeneracy parameter, with a factor of 2
accounting for spin and another factor of 2 for cone degeneracy; $v_{F}$ is
the Fermi velocity.

However, most materials naturally occur with charge doping where the Fermi level
or chemical potential $\mu$ is away from charge neutrality ($\mu=0$). 
When the graphene sheet is doped, the density-density correlation function
becomes more complicated. Following Ref.~\cite{PRB85_195427_2012},
the density-density correlation function on the imaginary frequency axis can be
written in terms of the dimensionless variables $\tilde{k}=k_{\parallel}/2k_{F}$
and $\tilde{\xi}=\hbar\xi/2E_F$ ($E_F=\hbar v_{F}k_{F}$ and
$k_{F}=\sqrt{4\pi n/g}$), as
\begin{equation}
\chi (\mathbf{k}, \xi) = - D_{0} \pr{1 + \frac{\tilde{k}^2}{4
\sqrt{\tilde{k}^{2} + \tilde{\xi}^{2}}} \pc{ \pi - f (\tilde{k}, \tilde{\xi})} }
\label{eq:chi}
\end{equation}
where $D_{0} = \sqrt{gn/(\pi \hbar^{2} v_{F}^{2})}$ is the density of
states at the Fermi level for doping concentration $n$. The function 
$f(\tilde{k}, \tilde{\xi})$ is defined as
\begin{gather}
f (\tilde{k}, \tilde{\xi}) = \arcsin \pc{\frac{1 - i \tilde{\xi}}{ \tilde{k}}} +
\arcsin \pc{\frac{1 + i \tilde{\xi}}{ \tilde{k}}}\nonumber \\
- \frac{i \tilde{\xi}-1}{\tilde{k}} \sqrt{1-\pc{ \frac{i \tilde{\xi}-1}{\tilde{k}}}^2}
+ \frac{i \tilde{\xi}+1}{\tilde{k}} \sqrt{1-\pc{ \frac{i \tilde{\xi}+1}{\tilde{k}}}^2}.
\end{gather}
In real graphene there are always deviations from the conical shapes of the band
structure and other than the lowest bands also contribute, thus the functions
used here are valid only for low frequencies ($\hbar \omega \lesssim 4$eV). 

\section{Atom near a single graphene sheet}
\label{sec:atomgraphene}

The simplest geometry is an atom at a distance $z_{A}$ away from a perfectly
flat graphene sheet. In case of undoped graphene, the relevant density-density
correlation function to be used in the reflection coefficients at a
graphene/vacuum interface $\mathrm{R}_{\mathrm{TM}}$ and 
$\mathrm{R}_{\mathrm{TE}}$ [Eqs.~\eqref{eq:rBo}], is the one shown in 
Eq.~\eqref{eq:xi_undoped}.

\begin{figure}[ht]
\begin{centering}
\includegraphics[width=8cm]{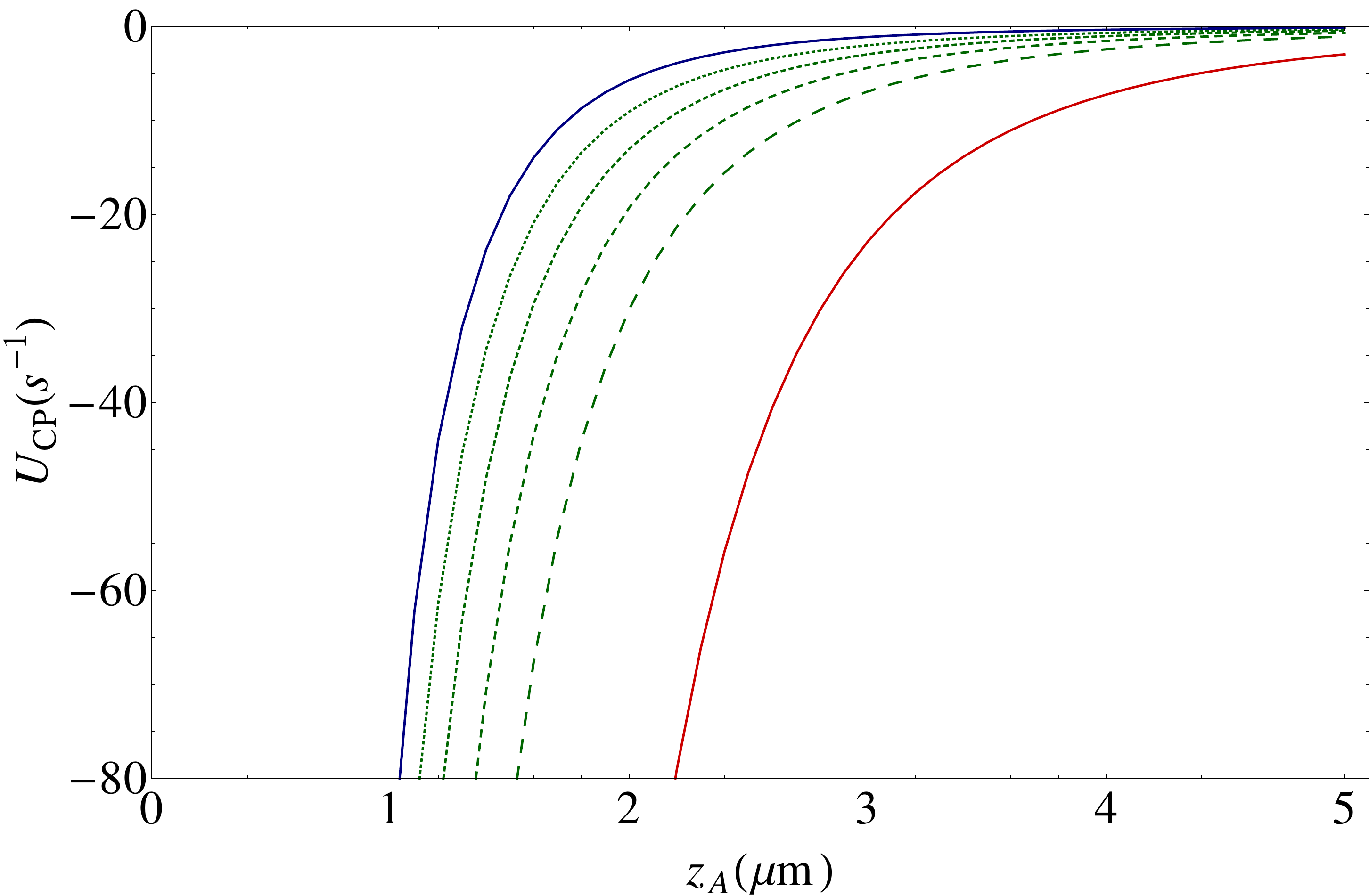}
\caption{(Color online) Casimir-Polder potential between a ground state rubidium atom and a
doped graphene sheet. The upper solid line (blue) is the result for undoped graphene,
while the dashed curves (green) are for doping densities $10^{10}$, $10^{11}$,
$10^{12}$ and $10^{13}$ cm$^{-2}$, respectively, from top to bottom. The solid bottom
line (red) is the result for a perfect conductor.
\label{results_doped_graphene}}
\end{centering}
\end{figure}

In the framework of this study, it is interesting to compare this result to the
interaction between an atom and a perfect conductor, where 
$\mathrm{R}_{\mathrm{TM}}=1$ and $\mathrm{R}_{\mathrm{TE}}=-1$. 
As an example, we have chosen a ground-state rubidium atom at zero temperature.
We found that the interaction between the atom and graphene is about $\sim 5\%$
of the interaction between an atom and a perfect conductor, see
Fig.~\ref{results_doped_graphene}. Using the same Dirac model for graphene, it
has already been shown that, at zero temperature, the
interaction between graphene and an ideal conductor is about $2.6\%$ of the
interaction between two perfect conductors separated by the same distance
\cite{PRB80_245406_2009}.

When the graphene sheets are doped, one has to use the reflection coefficients
given in Eq.~\eqref{eq:rBo} with $\chi (\mathbf{k},i\xi)$ defined by
Eq.~\eqref{eq:chi}. The results are shown in Fig.~\ref{results_doped_graphene}
and Table \ref{table:results}, where we present the results for doping
densities $10^{10}$, $10^{11}$, $10^{12}$ and $10^{13}$ cm$^{-2}$. 
\begin{table}
\caption{Casimir-Polder potential between rubidium
atoms in the ground state and graphene sheets at $z_A=1~\mu$m.
\label{table:results}}
\begin{center}
\begin{tabular}{l r}
  \hline
  \hline
  doping density (cm$^{-2})$ & $U_\mathrm{CP} (s^{-1})$ \\
  \hline
  \hline
  no doping & -90.987\\
  $10^{10}$ & -121.940\\
  $10^{11}$ & -165.489\\
  $10^{12}$ & -244.768\\
  $10^{13}$ & -371.140\\
  \hline
  \hline
\end{tabular}
\end{center}
\end{table}

\section{Graphene sheet above a gold substrate \label{sec:1sheet}}

\begin{figure}[ht]
\begin{centering}
 \includegraphics[width=7cm]{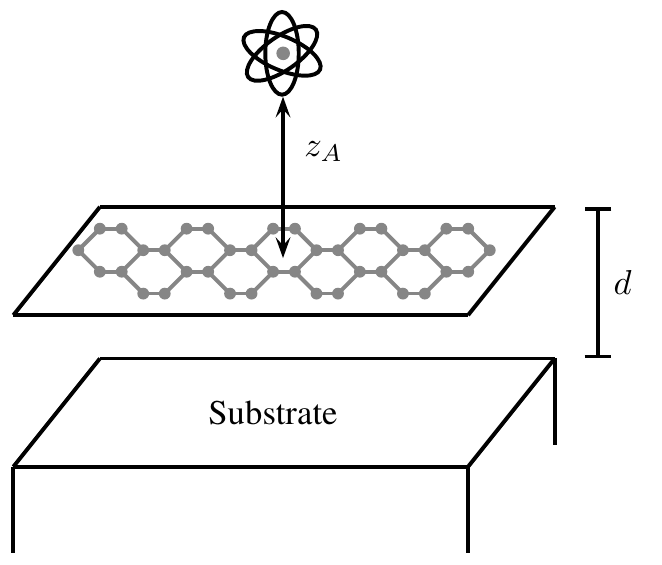}
 \caption{Scheme of an atom standing above a free standing graphene sheet above
a substrate.}
  \label{GoldGraphene_Atom}
\end{centering}
\end{figure}
The purpose of this manuscript is to show whether one (or more) freestanding
graphene sheet could be a candidate to shield the effects of a substrate
below on an atom above the graphene sheets (see Fig.~\ref{GoldGraphene_Atom}).
For a single graphene layer, the system will in effect be a layered medium of
the structure $1\nmid2\mid3$, where the graphene sheet (denoted by $\nmid$)
separates the free-space regions 1 and 2, the index 3 denotes the substrate
(subscript $s$) with permittivity $\varepsilon(\omega)$. The Fresnel
coefficients for this geometry can be written as \cite{ChewBook}
\begin{eqnarray}
\tilde{R} = R_{G} + \frac{T_{G} R_{0s} T_{G} e^{2 i k_{0 z} d}}{1 - R_{G} R_{0s}
e^{2 i k_{0z} d}},
\end{eqnarray}
where $T_{G}$ is the transmission coefficient through the graphene layer. 
For the different modes we will have different coefficients, such that $T^{\mathrm{TE}}_G = 1+R^{\mathrm{TE}}_G$ is valid only for the amplitude coefficient of the TE mode and the conditions $T^{\mathrm{TM}}_G = 1-R^{\mathrm{TM}}_G$ should be used for the TM mode. The Fresnel coefficients for this structure $1\nmid2\mid3$ should be written as
\begin{eqnarray}
\label{eq:R3}
\tilde{R}^{\mathrm{TE}} &= \frac{R^{\mathrm{TE}}_{G} + R^{\mathrm{TE}}_{0s} e^{2 i k_{0 z} d} + 2 R^{\mathrm{TE}}_{G} R^{\mathrm{TE}}_{0s} e^{2 i k_{0 z} d}}{1 - R^{\mathrm{TE}}_{G} R^{\mathrm{TE}}_{0s} e^{2 i k_{0z} d}}, \\
\tilde{R}^{\mathrm{TM}} &= \frac{R^{\mathrm{TM}}_{G} + R^{\mathrm{TM}}_{0s} e^{2 i k_{0 z} d} - 2 R^{\mathrm{TM}}_{G} R^{\mathrm{TM}}_{0s} e^{2 i k_{0 z} d}}{1 - R^{\mathrm{TM}}_{G} R^{\mathrm{TM}}_{0s} e^{2 i k_{0z} d}}.
\end{eqnarray}

The reflection coefficients for TE and TM waves at the interface between free
space and the substrate are the usual Fresnel coefficients
\begin{eqnarray}
 \mathrm{R}_{\mathrm{TM}}^{0s} &= \frac{\varepsilon_{s} \gamma_{0z} -
\gamma_{sz}}{\varepsilon_{s} \gamma_{0z} + \gamma_{sz}}, \\
 \mathrm{R}_{\mathrm{TE}}^{0s} &= \frac{\gamma_{0z} - \gamma_{sz}}{ \gamma_{0z}
+ \gamma_{sz}}.
\end{eqnarray}
In the following, we will present our results for the Casimir-Polder
interaction with both doped and undoped graphene sheets. As a substrate material
we have chosen gold, a material used in several experimental setups, whose
permittivity we describe by the Drude model
\begin{eqnarray}
\varepsilon (\omega) = 1 - \frac{\omega_{pe}^{2}}{\omega (\omega + i \gamma_{e})}
\end{eqnarray}
with parameters $\omega_{pe}=1.37\times 10^{16}\mbox{s}^{-1}$ and 
$\gamma_{e} = 4.12 \times 10^{13} \mbox{s}^{-1}$.

\subsection{Undoped graphene sheet}

In this simple geometry with only one graphene sheet, the total
Casimir-Polder potential of the graphene-substrate system is limited by the
potential of the single graphene sheet and that for the gold substrate. At
small distances $d$ between graphene and substrate the Casimir-Polder
interaction felt by the atom is dominated by the interaction with the gold
substrate. With increasing distance $d$, the Casimir-Polder potential is well
approximated by that of a single graphene sheet.
\begin{figure}[ht]
\begin{centering}
\includegraphics[width=8.5cm]{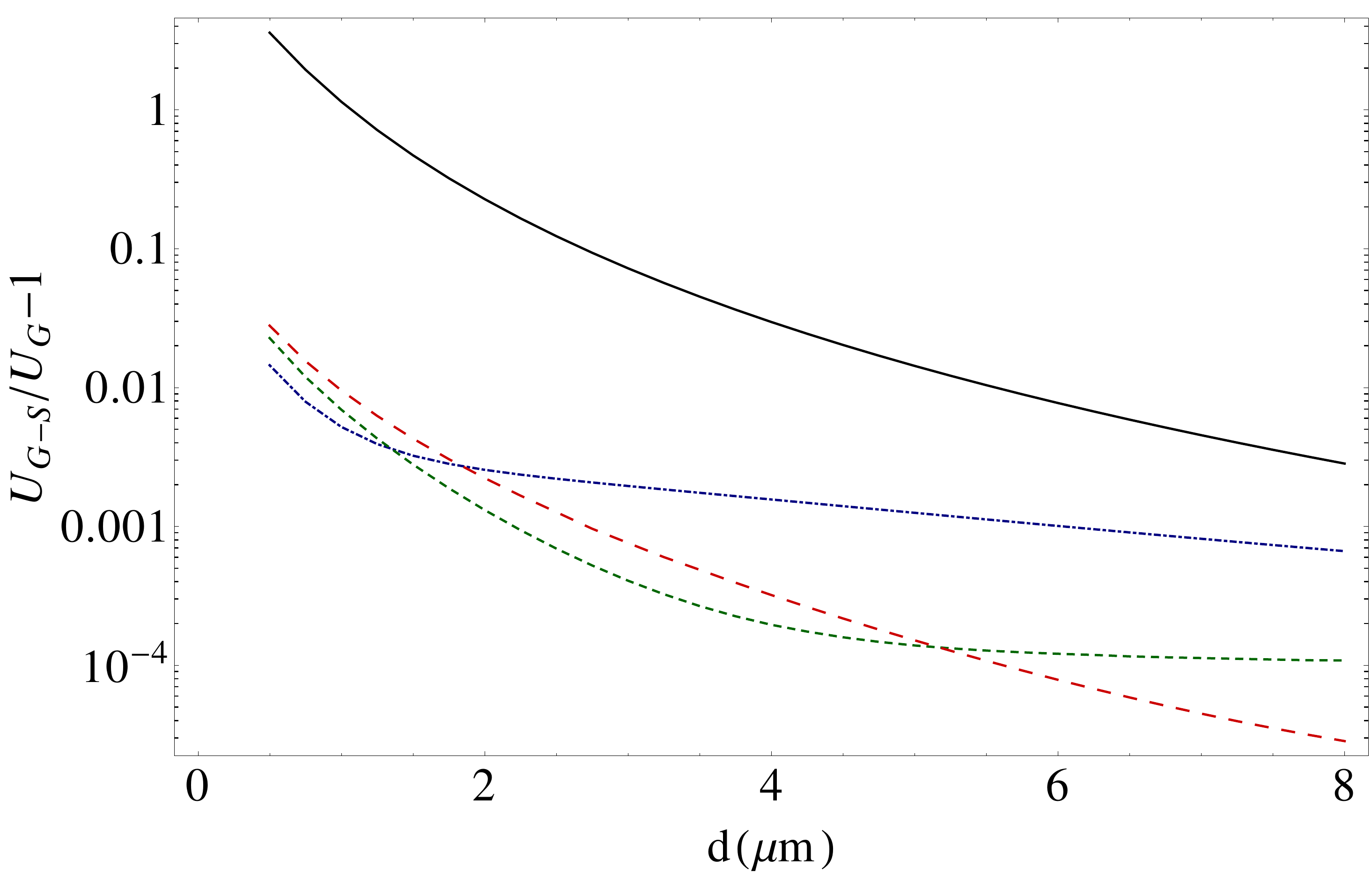}
\caption{(Color online) Normalized Casimir-Polder potential of a rubidium atom
in the ground state (black line), 32S (blue dotted line), 43S (green dashed
line) and  54S (red dashed line) at $z_{A}=1~\mu$m for different distances $d$
between the undoped graphene sheet and gold. \label{results_shield}}
\end{centering}
\end{figure}

In order to quantify the shielding effect of graphene, we fix the atom's
position at $z_{A} = 1~\mu$m, vary the distance $d$ between the graphene and the
substrate, and normalize these results to the Casimir-Polder potential without
the substrate at the same distance $z_{A}$.

Due to the recent experimental progress in working with atoms in Rydberg states,
one might look at the differences that may arise from having an atom in an
excited state rather than in its ground state. The atomic transition frequencies
of a highly excited Rydberg atom are in a window of frequencies in which
graphene absorbs well \cite{PRL108_047401_2012, nnano7_330_2012}, so that one
would expect a larger Casimir-Polder shift for the atom-graphene-gold system
than for the corresponding atom-gold system.
 
For the calculation of the interaction energy between an atom in a excited state
and a surface one has to add a resonant contribution to the usual non-resonant
Casimir-Polder potential \eqref{eq:Ucp_1}
\begin{gather}
U_{\mathrm{CP}}^{\mathrm{R}} \pc{z_{A}} = \frac{\mu_{0}}{4 \pi} \sum_{k \neq n}
\omega_{n k}^{2} \mathbf{d}_{n k} \cdot \mathbf{d}_{k n} \int_{0}^{\infty} d
\kappa_{0z} e^{-2 \kappa_{0z} z_{A}} \nonumber \\
\times  \pr{\mathrm{Re}\pr{\mathrm{R}_{\mathrm{TE}}} +
\mathrm{Re}\pr{\mathrm{R}_{\mathrm{TM}}} \pc{1 + \frac{2 \kappa_{0z}^{2}
c^{2}}{\omega^{2}}}}.  
\end{gather}
The Casimir-Polder potentials for a selection of Rydberg states of rubidium are
shown in Fig.~\ref{results_shield} and compared with the corresponding results
for the ground state. We can clearly see that for the excited states, the
shielding properties of graphene are highlighted. The differences in the potential experienced by the
various states reflect the resonances of the different atomic transitions
frequencies allowed for each state.

\subsection{Doped graphene sheet}

\begin{figure}[ht]
\begin{centering}
 \includegraphics[width=8.5cm]{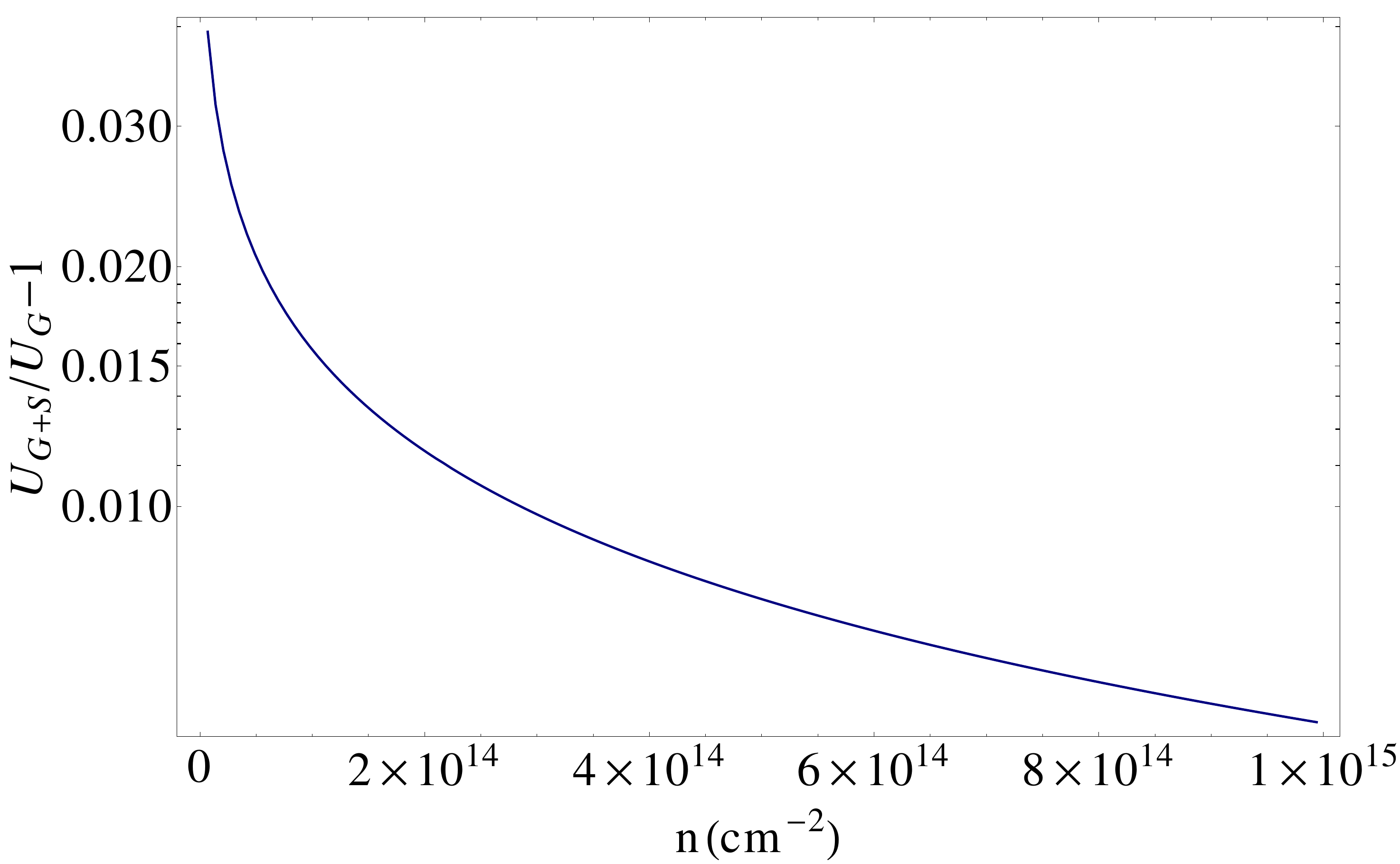}
\caption{(Color online) Normalized Casimir-Polder potential of a rubidium atom in 
the ground state at $z_{A}=1~\mu$m and $d = 2~\mu$m for different 
doping densities. \label{results_doping}}
\end{centering}
\end{figure}

For doped graphene one has to use the density-density correlation function 
Eq.~\eqref{eq:chi} in the reflection coefficients Eq.~\eqref{eq:rBo}.
In Fig.~\ref{results_doping} we show the results for different doping
concentrations for fixed atom-graphene and graphene-gold distances. One
observes that for higher doping concentration the shielding effect
of graphene becomes somewhat better than at lower concentrations. This
is due to the fact that the conductivity increases and therefore the graphene
sheet more and more resembles a perfect conductor, see
Fig.~\ref{results_doped_graphene}.

\begin{figure}[ht]
\begin{centering}
 \includegraphics[width=8.5cm]{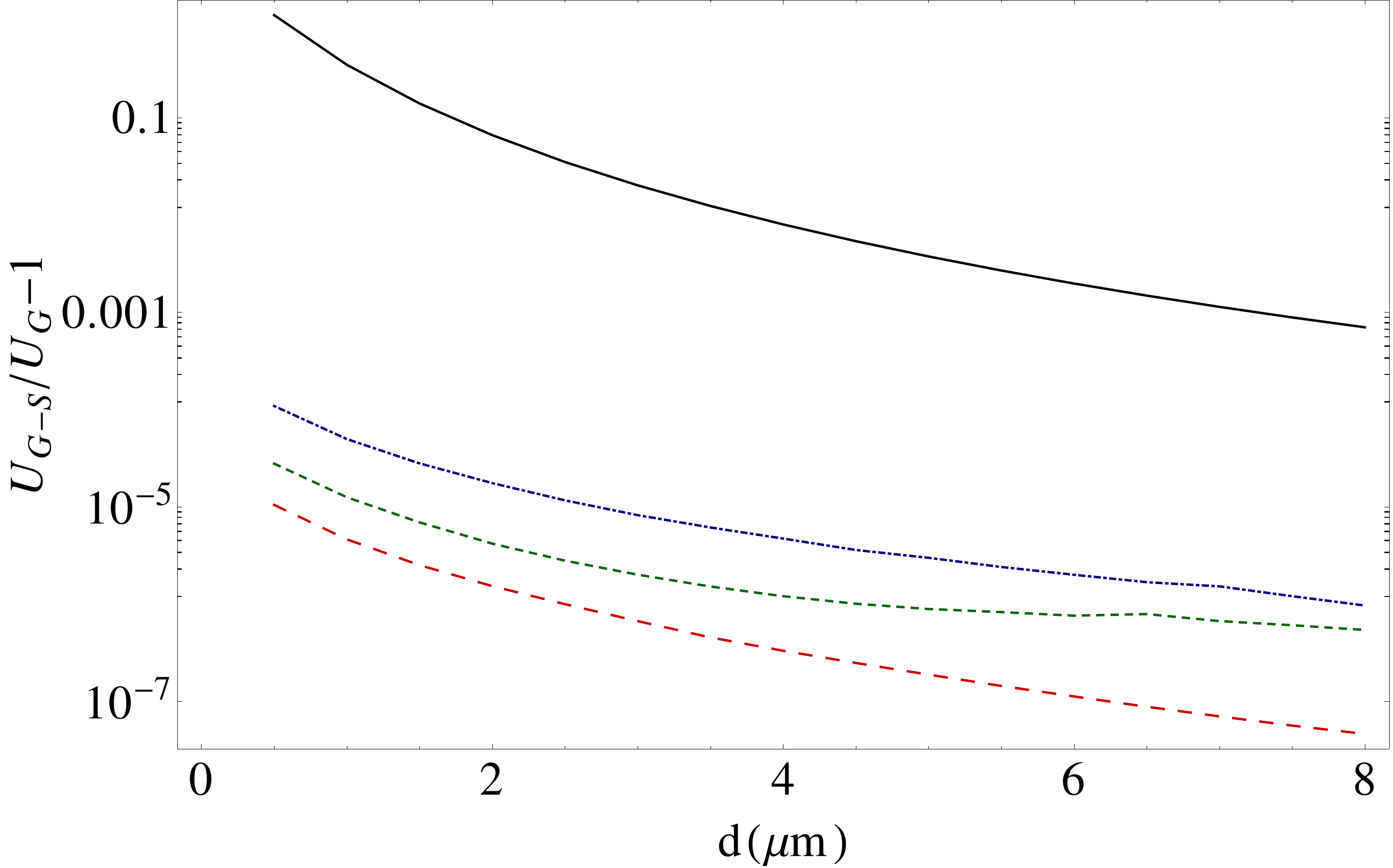}
\caption{(Color online) Normalized Casimir--Polder potential of a rubidium ($^{87}$Rb) atom in the ground  state (black line), 32S (blue dotted line), 43S (green dashed line) and  54S (red dashed line) at $z_{A}=1~\mu$m for different distances $d$ between doped graphene ($ n= 10^{12}$cm$^{2}$) and gold. 
\label{results_doped_states}}
\end{centering}
\end{figure}

Note that these results change for different atomic eigenstates, the
effect of which are shown in Fig.~\ref{results_doped_states} for ground and
excited state (the curves were calculated for a doping concentration of $ n=
10^{12}$cm$^{2}$). Each atom has different frequency transitions that influence
the strength of the atom-graphene coupling. In the same way, different doping
concentrations will also influence the absorbance of graphene, 
Fig.~\ref{results_doped_graphene}, so it is expected that each concentration 
and each atomic state to yield unique results.

\section{Bilayer graphene above a substrate} \label{sec:bilayer}

During the manufacturing of graphene, layers of varying thickness are typically
generated. Besides the 'pure' form of single-layer graphene, bilayers of two
weakly bound sheets are common. The natural form for bilayer graphene is the
AB-stacking, which is the basis of graphite. However, alternative stackings are
also available where one layer is rotated by some angle relative to the other
\cite{PRB86_075439_2012}. Here we focus on AB-stacking in which
half the atoms are aligned on top of one another whereas the other half are
located above the center of the hexagonal lattice of the opposite layer.

The conductivity of AB-stacked bilayer graphene can be found in
Refs.~\cite{PRB77_155409_2008,PRB86_075439_2012} for both doped and undoped
cases. The longitudinal conductivity for undoped AB-stacking ($\mu=0$) at zero
temperature can be written as
\begin{gather}
\sigma_{xx} (\omega) = \frac{e^2}{2 \hbar}  \nonumber \\
\times\bigg[ \bigg( \frac{\omega +2 \gamma}{2(\omega+\gamma)}+\frac{\omega -2
\gamma}{2(\omega-\gamma)} \Theta (\omega-2\gamma) \bigg) \Theta (\omega)
\nonumber \\
+  \frac{\gamma^{2}}{2 \omega^2} \pr{\Theta (\omega-\gamma) + \Theta
(\omega+\gamma)} \Theta (\omega-\gamma) \bigg] \,.
\end{gather}
and the perpendicular conductivity is given by 
\begin{gather}
\sigma_{zz} (\omega) = \frac{e^2}{4 \hbar} \pc{\frac{\gamma d}{\hbar v_F }}^2
\nonumber \\ \times  \bigg[
\frac{\omega}{2(\omega+\gamma)}+\frac{\omega}{2(\omega-\gamma)}
\Theta (\omega-2\gamma) \bigg] \Theta (\omega) 
\end{gather}
where $\Theta (x)$ is the Heaviside step function, $\gamma=0.4$~eV is the
interlayer hopping energy and $d=3.3$~\AA ~the interlayer distance.
These results for conductivity do not include the spin-orbit coupling. That
effect could be also considered to calculate the conductivity in order to cover
other possible effects \cite{PRB86_075439_2012}.

\begin{figure}[ht]
\begin{centering}
 \includegraphics[width=8.5cm]{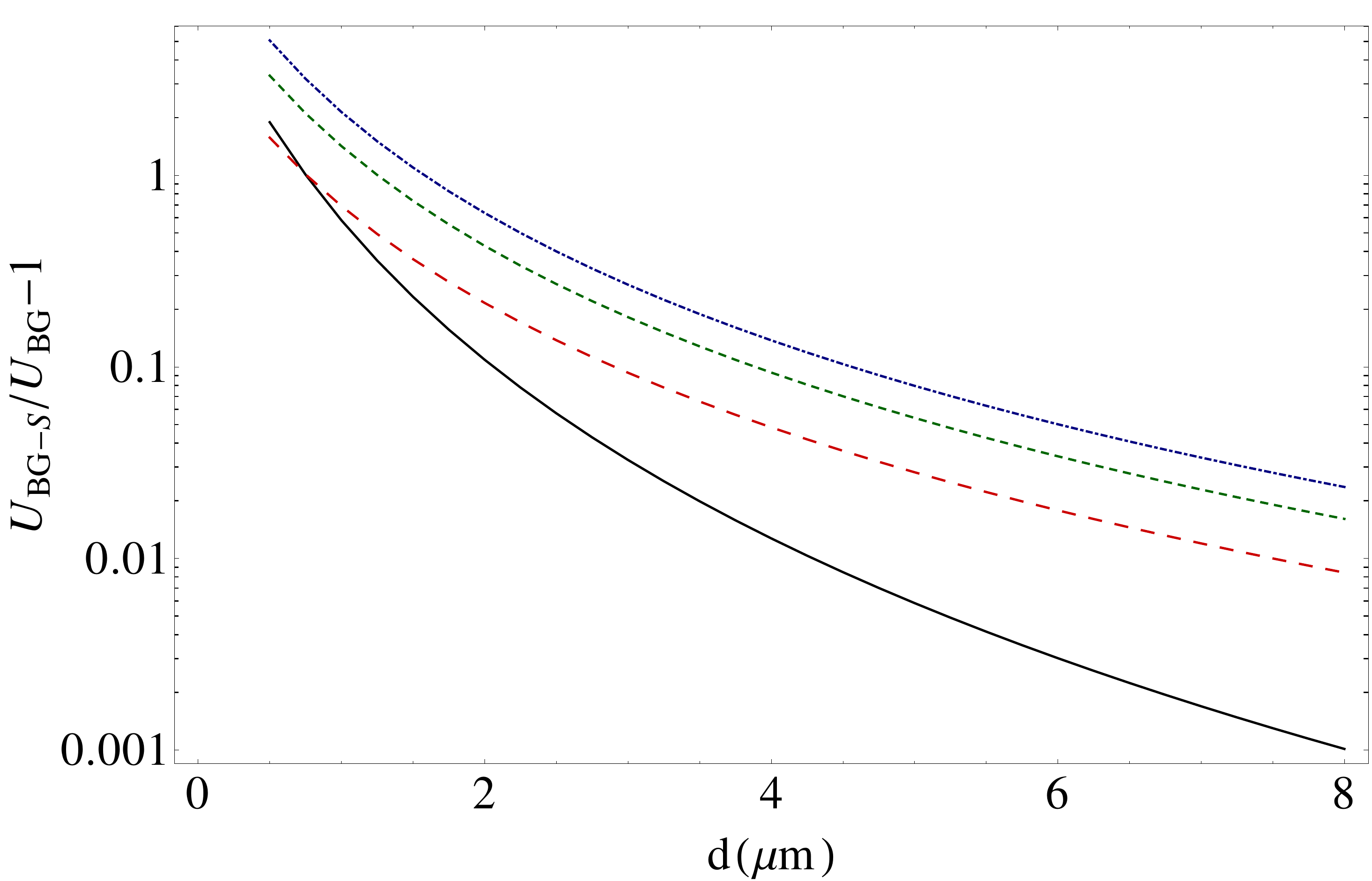}
\caption{(Color online) Normalized Casimir-Polder potential of a rubidium atom in the ground
state (black line), 32S (blue dotted line), 43S (green dashed line) and 54S (red
dashed line), at $z_{A}=1~\mu$m for different distances $d$ between the graphene
bilayer and gold.}
 \label{results_bilayer}
\end{centering}
\end{figure}

The conductivity at imaginary frequencies as required for the nonresonant
Casimir-Polder potential can be obtained from the Kramers-Kronig relation
\begin{gather}
\sigma (i \xi) = \frac{2}{\pi} \int_{0}^{\infty} d \omega \, \frac{\omega \,
\mathrm{Im} \sigma (\omega)}{\omega^2+ \xi^2}.
\end{gather}
In Fig.~\ref{results_bilayer} we show the Casimir-Polder potential of an atom
(either in its ground state or in a Rydberg state) next to a graphene bilayer.
When compared to a single graphene sheet, a bilayer of graphene does not
provide a better shielding for a ground state atom and for an excited one the
results are even more unfavourable, see Fig.~\ref{results_shield}.

\section{Conclusions \label{sec:conclusions}}

The knowledge of how to control and manipulate graphene systems opens the
possibility of a number of novel research possibilities. A layered structure
made from graphene could be used as an effective shield for the effects of a
substrate beneath, by patterning the graphene into disks and ribbons, one would
be able to create tunable filters, where the disks and nanoribbons would shield
the substrate and the void regions would only feel the effects of the
substrate. 

We have shown that, in some situations, in a atom-graphene-substrate system
the Casimir-Polder potential is dominated by the graphene potential.
For graphene-substrate distances larger than 4~$\mu$m we verified that for all
studied systems, one or more graphene membranes shield the effect of 
electromagnetic vacuum fluctuations emanating from a substrate.
The optical absorption of graphene is dominated by intraband transitions 
in the far-infrared spectral range and by interband transitions from 
mid-infrared to ultraviolet \cite{SolidStateCom152_1341_2012}. The coupling 
of graphene with different states will lead to different couplings as we could 
verify from the results obtained from ground and highly excited states. 
Each state has different frequency transitions which will influence the 
strength of the atom-graphene coupling.

These results show that shielding the Casimir-Polder forces is a rather
delicate issue. It has been proposed that, since hydrogen-switchable mirrors
are shiny metals which become optically transparent upon exposure to hydrogen,
the Casimir force between them should be stronger in air than in hydrogen.
However, that was been shown in Ref.~\cite{PNAS101_12_2004}
not to be the case, the reason being that, although the mirrors are indeed shiny
metals in air, this change in optical properties only affects the optical range
of frequencies. In order to have an effective change in the Casimir interaction
one would need to have the mirror which is strongly reflecting at all
frequencies \cite{NJOP8_235_2006}. 

A similar situation is encountered with graphene. The results for patterned
graphene in Ref.~\cite{PRL108_047401_2012} show that it is be possible to create
a system that is highly absorbing in a small frequency band, so as to tune
out the resonant part of the Casimir-Polder interaction for an atom in the
excited state. However, in order to be able to shield the non-resonant part
one would have to have a material that is more broadband absorbing.

Several factors could still be included to make our model more realistic.
Amongst them are finite temperature, corrugation of the freestanding
graphene sample and the presence of impurities. However, for clean enough
samples, these factors are normally considered as perturbations, not changing
the essentials of the Dirac model \cite{PRB80_245406_2009}.

We would like to  acknowledge fruitful discussions with E.A.~Hinds and S.Y.~Buhmann.
SR is supported by the PhD grant SFRH/BD/62377/2009 from FCT, co-financed by
FSE, POPH/QREN and EU.
\bibliography{phdthesis}
\end{document}